\documentclass[noshowpacs,amsmath,
twocolumn,
superscriptaddress,
8pt,aps,prb]{revtex4-1}
\usepackage{setspace}
\usepackage{amsmath}
\usepackage{bm}
\usepackage{graphicx}
\usepackage[nearskip,margin = 0pt]{subfig}

\usepackage{verbatim}
\usepackage{amsfonts}
\usepackage{amssymb}
\usepackage{upgreek}
\usepackage[colorlinks,linkcolor=blue,anchorcolor=blue,citecolor=blue,urlcolor=black]{hyperref}
\usepackage{epstopdf}
\usepackage{xcolor}
\usepackage{booktabs}
\usepackage{tabularx}
\usepackage{xtab}
\usepackage{ragged2e}
\DeclareGraphicsExtensions{.pdf,.eps,.png,.jpg,.mps} 

\begin{document}
\title{Gain-assisted chiral soliton microcombs}

\author{Teng Tan$^{1,\#}$, Hao-Jing Chen$^{2,\#}$, Zhongye Yuan$^{1,\#}$, Yan Yu$^{2,\#}$, Qi-Tao Cao$^{2}$, Ning An$^{1}$, Qihuang Gong$^{2}$, Chee Wei Wong$^{3,*}$, Yunjiang Rao$^{1,4,*}$, Yun-Feng Xiao$^{2,*}$, Baicheng Yao$^{1,*}$\\
\vspace{3pt}
$^1$Key Laboratory of Optical Fiber Sensing and Communications (Education Ministry of China), University of Electronic Science and Technology of China, Chengdu 611731, China.\\
$^2$State Key Laboratory for Mesoscopic Physics and Frontiers Science Center for Nano-optoelectronics, School of Physics, Peking University, Beijing 100871, China.\\
$^3$Fang Lu Mesoscopic Optics and Quantum Electronics Laboratory, University of California, Los Angeles, CA 90095, United States.\\
$^4$Research Centre of Optical Fiber Sensing, Zhejiang Laboratory, Hangzhou 310000, China.\\
$^{\#}$These authors contributed equally to this work.\\
$^{*}$Corresponding authors: yaobaicheng@uestc.edu.cn; yfxiao@pku.edu.cn; yjrao@uestc.edu.cn;
cheewei.wong@ucla.edu.}

\date{\today}

\maketitle

\noindent\textbf{The emerging microresonator-based frequency combs\cite{del2007optical,kippenberg2011microresonator,levy2010cmos,razzari2010cmos,jung2013optical,he2019self,fujii2020dispersion} revolutionize a broad range of applications from optical communications to astronomical calibration\cite{Marin2017Microresonator,corcoran2020ultra,suh2019searching,obrzud2019microphotonic}.
Despite of their significant merits, low energy efficiency and the lack of all-optical dynamical control severely hinder the transfer of microcomb system to real-world applications\cite{bao2019laser}.
Here, by introducing active lasing medium into the soliton microcomb, for the first time, we experimentally achieve the chiral soliton with agile on-off switch and tunable dual-comb generation in a packaged microresonator. It is found that such a microresonator enables a soliton slingshot effect, the rapid soliton formation arising from the extra energy accumulation induced by inter-modal couplings. Moreover, tuning the erbium gain can generate versatile multi-soliton states, and extend the soliton operation window to a remarkable range over 18 GHz detuning. Finally, the gain-assisted chirality of counter-propagating soliton is demonstrated, which enables an unprecedented fast on-off switching of soliton microcombs. The non-trivial chiral soliton formation with active controllability inspires new paradigms of miniature optical frequency combs and brings the fast tunable soliton tools within reach.}

\begin{figure*}[hbtp]
\centering
\captionsetup{singlelinecheck=no, justification = RaggedRight, labelfont={bf},name={Fig.}}
\includegraphics[width=16cm]{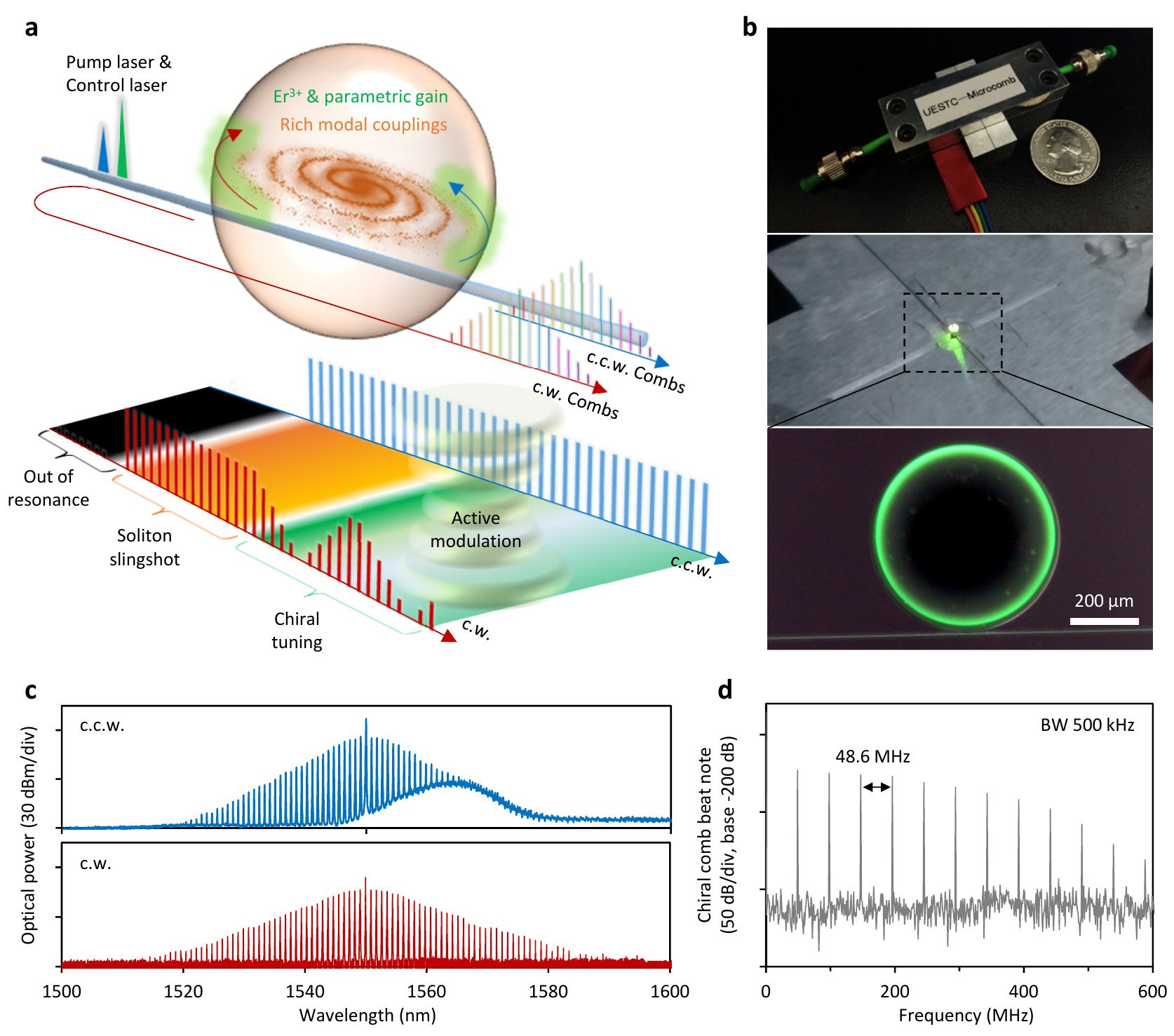}
\caption{\textbf{Schematic of chiral soliton slingshot. a,} Architecture of the soliton slingshot effect and active chiral tuning. A 980 nm control laser and a 1550 nm pump laser are coupled into an erbium doped silica microsphere through a tapered fiber. Soliton comb bursts emerge in both c.w. and c.c.w. direction (orange region in bottom panel), chirality of soliton comb output is controlled by the assistance of active modulation. (green region in bottom panel). 
\textbf{b,} Images of a packaged microcomb device. The zoom-in microscopic picture illustrates the erbium-based excitation. \textbf{c,} Measured optical spectra of the counter-propagating soliton combs. 
\textbf{d,} The RF beat note between the c.w. and c.c.w. soliton combs, where the frequency spacing of the peaks are 48.6 MHz.}
\label{pic:Fig1}
\end{figure*}

\begin{figure*}[ht]
\centering
\captionsetup{singlelinecheck=no, justification = RaggedRight,labelfont={bf},name={Fig.}}
\includegraphics[width=16cm]{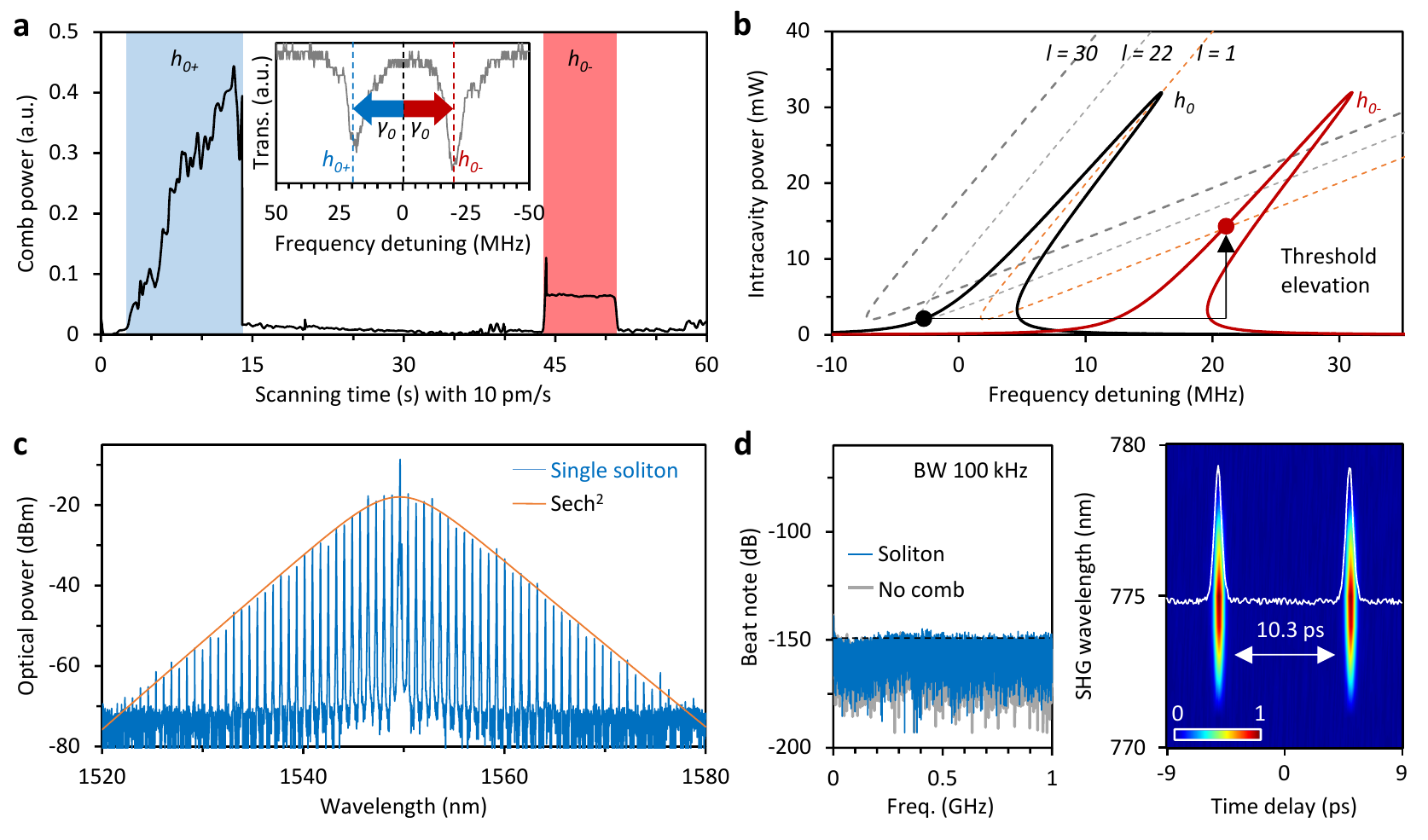}
\caption{\textbf{Single soliton slingshot. a,} The comb power evolution during the pump frequency scanning over the hybrid mode  $h_{0+} \& h_{0-}$ with 10 pm/s scanning speed. Inset: spectrum of hybrid modes induced by modal coupling, where $h_{0+(-)}$ denotes the high-frequency (low-frequency) hybrid mode. 
\textbf{b,}  Schematic for parametric oscillation threshold elevation. The black (red) curve shows the intracavity power as a function of cavity–pump frequency offset without (with) modal-coupling-induced red shift. Dashed lines are parametric oscillation stability curve of 1, 22, 30-FSR sideband.
\textbf{c,} The measured optical spectrum of the slingshot soliton comb with a fixed pump frequency, of which the envelope is fitted by a $\mathrm{sech}^2$ function.
\textbf{d,} Left panel: the RF spectrum of the obtained comb signal in \textbf{c}, where the resolution-bandwidth is set to be 100 Hz.
Right panel: FROG trace displays the signal of the single soliton with period of 10.3 ps.}
\label{pic:Fig2}
\end{figure*}

\begin{figure*}[!ht]
\centering
\captionsetup{singlelinecheck=no, justification = RaggedRight,labelfont={bf},name={Fig.}}
\includegraphics[width=16cm]{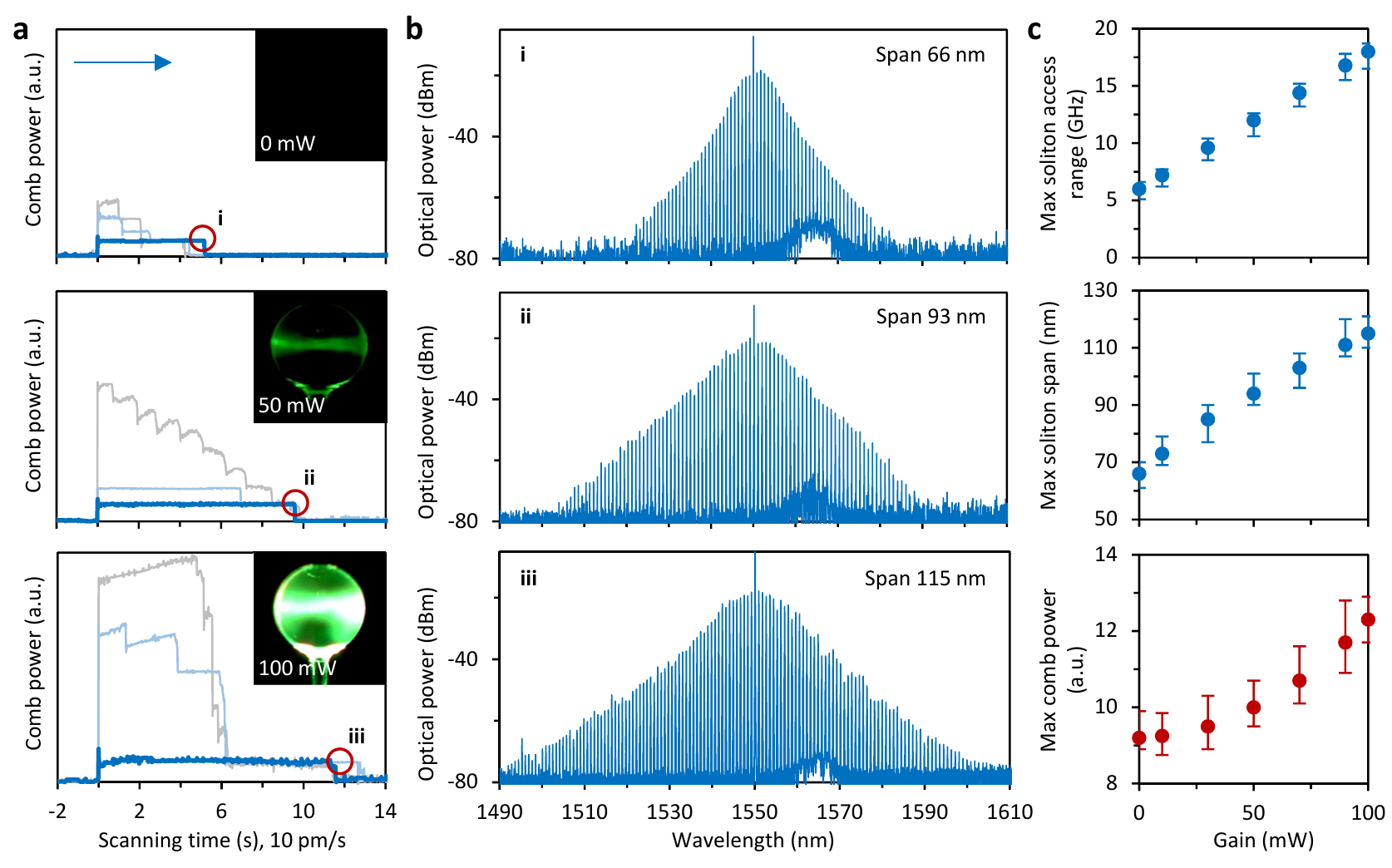}
\caption{\textbf{Soliton slingshots regulated by optical gain. a,} Measured comb power evolution traces under control laser power of 0 mW, 50 mW and 100 mW with a scanning speed of 10 pm/s. The laser tuning direction is denoted by the blue arrow. The maximum soliton access range reaches 6.2 GHz, 12.5 GHz, and 18 GHz. Insets: Microscopic images of the active microsphere. 
\textbf{b,} Optical spectra of single slingshot soliton at maximum detuning under the control laser power of 0 mW (i), 50 mW (ii) and 100 mW (iii). The corresponding spectral spans are 66 nm, 93 nm and 115 nm.
\textbf{c,} Gain enhanced soliton formation. From top to bottom: the maximum soliton access range, the maximum soliton span,  and the soliton power, as increasing the power of control laser from 0 to 100 mW.}
\label{pic:Fig3}
\end{figure*}

\begin{figure*}[!ht]
\centering
\captionsetup{singlelinecheck=no, justification = RaggedRight,labelfont={bf},name={Fig.}}
\includegraphics[width=16cm]{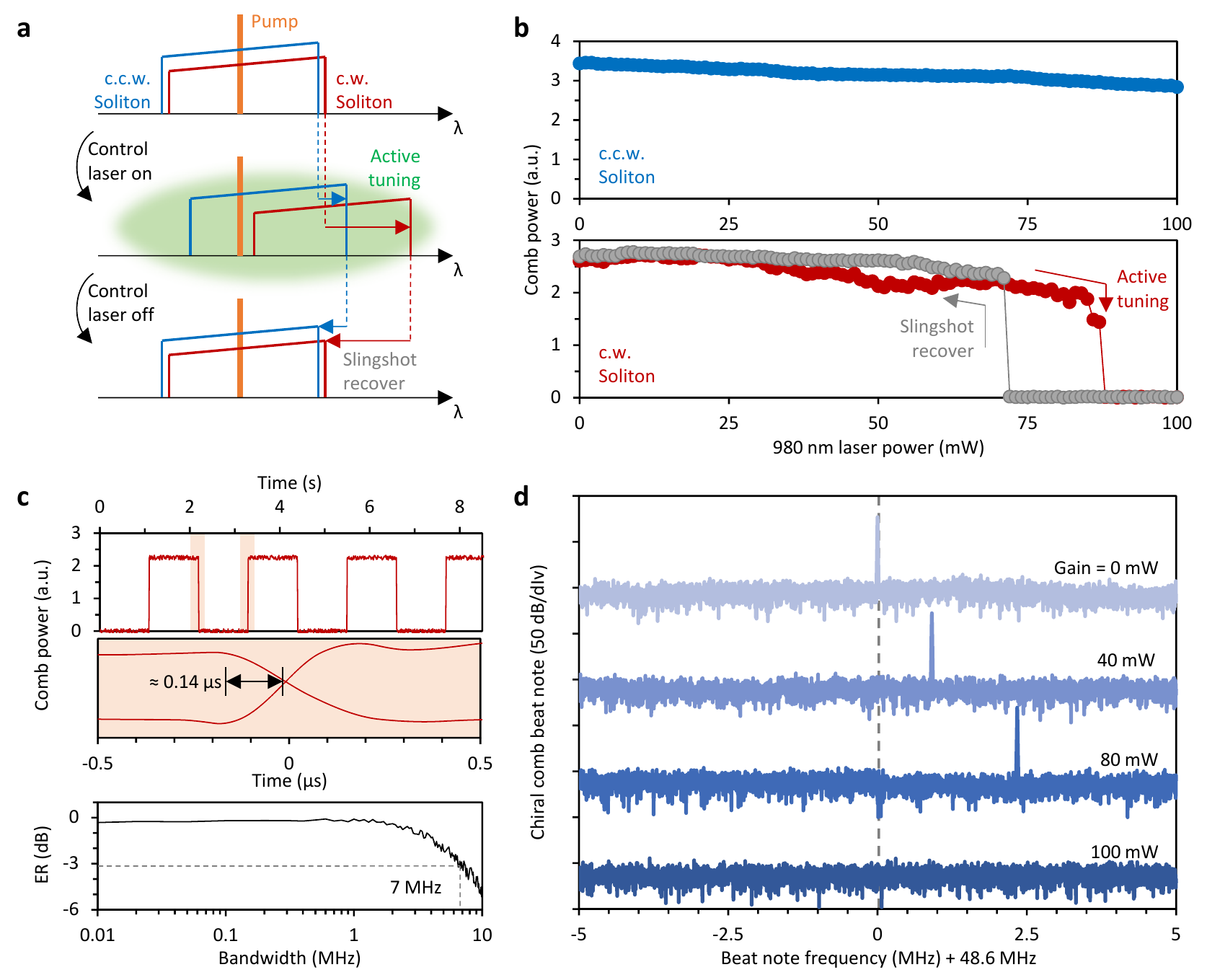}
\caption{\textbf{On-off switch of soliton microcombs. a,} Principle of on-off chirality control of the counter-propagating soliton. 
The blue and red frame denote c.c.w and c.w. soliton existence (detuning) range. The control laser regulates the refractive index of c.w. and c.c.w. mode family asymmetrically, resulting in the chiral modulation of the mode effective detuning. 
\textbf{b,} Measured comb power trace via tuning the control laser power. The c.w. soliton experiences the rapid annihilation via active tuning (red) and recovery via the slingshot effect, while the c.c.w comb power (blue) remains almost unchanged.  
\textbf{c,} Output power of the soliton comb in c.w. direction by modulating the control laser power, showing a on-off delay time $\approx$ 0.14  $\upmu$s, corresponding to a 3-dB extinction ratio (ER) modulation bandwidth of 7 MHz.
\textbf{d,} Measured beat note of the counter-propagating soliton (the frequency offset is 48.6 MHz), where the beat frequency increases with the control laser power.} 
\label{pic:Fig4}
\end{figure*}

\vspace{6pt}
\noindent
Temporal solitons in optical microresonators\cite{herr2014temporal,obrzud2017temporal}, offering discrete pulses and equally spaced frequency rulers, play a pivotal role in precise measurement applications\cite{xue2015mode,huang2015mode,kippenberg2018dissipative,jang2020nanometric,huang2016broadband}. So far, frequency microcombs have been applied widely in laser-based light detection and ranging (LiDAR) \cite{trocha2018ultrafast,suh2018soliton,riemensberger2020massively}, high-speed optical communication networks\cite{Marin2017Microresonator,corcoran2020ultra,geng2018terabit}, on-chip spectroscopy \cite{suh2016microresonator,stern2020direct}, and quantum sources \cite{reimer2016generation,lodahl2017chiral,kues2019quantum}. 
To push the practicality of these applications, considerable effort has been made recently, such as turnkey soliton generation \cite{shen2020integrated}, broadband spectral coverage \cite{li2017stably,chen2020chaos,yang2020coherent}, high efficiency energy conversion \cite{xue2017microresonator,chang2020ultra,stern2018battery}. Beyond the significant progresses in soliton generation, the ability to dynamically observe and control the soliton properties\cite{li2020real} also lies at the heart of various high-performance applications. Recently, more and more approaches to the soliton regulation are pursued, involving piezoelectric \cite{niu2018repetition,liu2020monolithic}, gate-tunable\cite{yao2018gate} and thermal control\cite{dutt2018chip,drake2020thermal}, for the modulation of repetition rate, dispersion and cavity-pump detuning.

Despite these enormous advances, the existing strategies for the dynamical control of soliton microcomb are restricted to passive microresonators, while the optical gain provides a potential way to enable all-optical control of the microcomb, an exotic degree of freedom that is exclusive to active microresonators \cite{villegas2014all,zhu2018all,chen2017exceptional}.
In this work, we report the soliton slingshot effect and the chiral modulation of the counter-propagating soliton in an erbium-doped active microcavity. The soliton slingshot effect arises from the extra energy accumulation induced by the modal coupling in the over-moded resonator. The operation window of the slingshot single soliton can be actively regulated by tuning the erbium gain, with a maximum of 18 GHz. Furthermore, the chirality of the counter-propagating soliton is demonstrated and dynamically controlled through the erbium gain, realizing a rapid on-off switch of soliton with a modulated bandwidth of 7 MHz.

In the experiment, a silica microsphere with a diameter $\approx$ 600 $\upmu$m and 99.7 GHz free spectral range (FSR) is prepared (Fig. \ref{pic:Fig1}a), which contains rich mode families, for example 241 transverse modes per FSR (see Supplementary Information).
The erbium particles are then doped on the cavity surface via the arc plasticity annealing technique \cite{kotz2017three} (details see Methods and Supplementary Information), which is characterized by the green light emission under the pump of a 980 nm control laser (Fig. \ref{pic:Fig1}b). The device coupled with a fiber taper is mounted into an all-fiber-connected centimeter-size package (Fig. \ref{pic:Fig1}b), for facilitating measurements and enabling portability. By tuning the 1550 nm pump laser frequency into the cavity resonance, soliton combs are obtained, exhibiting a smooth sech$^2$ spectral envelope, as shown in Fig. \ref{pic:Fig1}c. 

Different from the conventional generation process, the solitons directly burst in an extremely short time, behaving as a slingshot, in which extra optical energy is accumulated in the pump mode and then transfered to the side modes. Moreover, the soliton combs are generated in both counterclockwise (c.c.w.) and clockwise (c.w.) direction due to the backscattering of the pump laser induced by the cavity-fiber coupling spot (see Supplementary Information). Here the backscattering excites different transverse modes in c.w. direction, enabling the backward soliton formation in another mode family. The dynamics of multimode co-existence in the microresonator is demonstrated by real-time video recording (see Supplementary Movie 1). The counter-propagating solitons can be also formed in the same mode family (Extended Data Fig. \ref{pic:ED1}). The counter-propagating solitons share the same pump wavelength, but are different in both the spectral shape and bandwidth (Fig. \ref{pic:Fig1}c). Besides, by injecting both the c.c.w. and c.w. soliton combs into a photodetector, the radio-frequency (RF) beat note signal is obtained through a electrical spectrum analyzer, indicating a 48.6 MHz repetition rate difference of the counter-propagating solitons (Fig. \ref{pic:Fig1}d). Such dual-comb in opposite directions can also turn into chiral regime with the unbalanced intensity output by adjusting the intracavity erbium-based gain, as shown in Fig. \ref{pic:Fig1}a. 

The aforementioned soliton slingshot phenomenon is attributed to the modal-coupling-induced energy accumulation. In a microsphere resonator with a few hundreds of microns in diameter, numerous transverse modal couplings emerge (see Supplementary Information), forming hybrid modes $h_{0+}$ and $h_{0-}$ with a considerable frequency shift $\pm \gamma_0$ (Inset of Fig. \ref{pic:Fig2}a). 
When the frequency shift $\gamma_0$ is large enough, the parametric oscillation threshold of the red-shifted mode $h_{0-}$ increases along the oscillation stability curve  (Fig. \ref{pic:Fig2}b), giving rise to the extra accumulation of intracavity pump power (see Supplementary Information for details). Once the intracavity power reaches the elevated threshold, a rapid energy transition from the pump mode to the side modes takes place as well as the subsequent burst of soliton microcombs. As for the blue-shifted mode $h_{0+}$, the oscillation threshold remains unchanged, so that no extra energy accumulation of the pump mode is provided in this resonance. The soliton slingshot processes are recorded by an optical spectrum analyzer in Supplementary Movie 2.

Experimentally, as the pump laser frequency scans over the two modes from the blue to the red side with a speed of 10 pm/s, the output comb power evolution trace is monitored in real time. It is found that the high-frequency mode $h_{0+}$ experiences a regular comb generation, while in the $h_{0-}$ resonance the soliton slingshot effect is observed, featuring an abrupt rising step at 44s (Fig. \ref{pic:Fig2}a, detail evolution process in spectral and time domain see Extended Data Fig. \ref{pic:ED2}). The optical spectrum of the rapidly generated soliton at $h_{0-}$ mode spans from 1500 nm to 1600 nm, meeting the sech$^2$ function fitting of the soliton spectral envelope (orange dash line in Fig. \ref{pic:Fig2}c) and a repetition rate of $\approx$ 99.7 GHz. The corresponding low-frequency spectrum at 0 to 1 GHz in Fig. \ref{pic:Fig2}d reveals the phase-locked property of the microcombs. The temporal profile is also measured by frequency-resolved optical gating (FROG), indicating the narrow pulsewidth of the slingshot soliton being $\approx$ 350 fs, and pulse-to-pulse distance of 10.3 ps. Moreover, due to the rich modal couplings in the cavity, such single soliton slingshots are generated with a high possibility. By repeating 100 times scan for each of the 40 microsphere samples, we obtain 3,640 cases of soliton slingshot ($>$90\% in possibility)in which 2,575 cases generates the single soliton state ($>$60\% in possibility, details in Section 2 in Supplementary Information). We would note that the similar spontaneous formation process of the soliton pulse is also reported recently \cite{yu2020spontaneous,zhou2019soliton}.

The demonstrated soliton microcomb slingshot can be regulated actively by tuning the optical gain from the doped erbium particles. 
By launching a 980 nm control laser into the cavity via the same tapered fiber, the erbium particles are excited, characterized by the down-conversion fluorescence (see the inset of Fig. \ref{pic:Fig3}a). With increasing the control laser power, the erbium gain rises considerably, compensating significantly the photon losses. This energy enhancement further leads to the extension of soliton existence range as well as greater possibilities of multi-soliton and soliton crystal slingshots.

By repeating the pump laser frequency scan from blue to red detuned region under control laser power of 0 mW, 50 mW and 100 mW, we record the comb power evolution traces, as shown in Fig. \ref{pic:Fig3}a. The extension of slingshot soliton existence range and the generation of multi-soliton state with higher optical power are obtained, as increasing the erbium gain. Details of obtained multi-soliton and soliton crystal are presented in Extended Data Figure \ref{pic:ED3}. The slingshot single soliton existence ranges are 6.2 GHz, 12.5 GHz and 18 GHz under control laser power of 0 mW, 50 mW and 100 mW, respectively. Such an extended single soliton existence range also increases the maximum spectral span of the soliton microcomb, corresponding to 66 nm, 93 nm and 115 nm span under control laser power of 0 mW (i), 50 mW (ii) and 100 mW (iii) (Fig. \ref{pic:Fig3}b). We summarize such gain based enhancements as increasing the power of control laser in Fig. \ref{pic:Fig3}c.

To exploit the controllability of this slingshot soliton in the active system, we demonstrate the fast on-off switch of soliton microcombs. As aforementioned, the 1550 nm pump light is partially backscattered to c.w. direction at the cavity-fiber coupling point and excites different transverse mode families in c.w. and c.c.w. directions. Thus, the counter-propagating solitons are generated in the two disparate mode families, while sharing the same pump frequency. The refractive index of c.c.w. and c.w. mode families reacts differently to the erbium excitation, thus resulting in the resonance shift with different velocities in opposite directions.   
The c.w. resonance frequency, with higher sensitivity to the erbium-based modulation, is tuned out of soliton existence range when increasing the control laser power. By decreasing the control laser power, the c.w. resonance is tuned back towards shorter wavelengths, resulting in the slingshot recovery of c.w. soliton microcomb. The less sensitive c.c.w. soliton remains almost unaffected in the entire modulation process, enabling the on-off chirality control of counter-propagating soliton (More details are shown in the Extended Data Fig. \ref{pic:ED4} and Supplementary Section 2). Since the slingshot effect requires energy accumulation in the pump mode, the c.w. comb power exhibits hysteresis behavior, because the iterated-slingshot of soliton needs higher optical energy accumulation than the original soliton build-up. Meanwhile, the c.c.w. comb power remains almost unchanged, as shown in Fig. \ref{pic:Fig4}b. The fact that the counter-propagating solitons belong to different mode families is also proven by the beat note signal between the output c.c.w. and c.w. microcombs, forming a set of dual-comb with different repetition rate. Moreover, the asymmetrical regulation of counter-propagating soliton provides a novel method to tune the repetition rates of the dual comb actively. By tuning the control laser power from 0 to 80 mW, the repetition rate difference of the dual comb changes from 48.6 MHz to 51.2 MHz (Fig. \ref{pic:Fig4}d). It is noted that when the control laser power reaches 100 mW, the c.w. soliton vanish gradually, and no dual-comb beating is observed. Due to the instantaneous response of all-optical control, the fast on-off switch is realized. By turning a 100 mW control laser on and off periodically, the output intensity of the c.w. soliton microcombs is well modulated in a square wave, with modulation extinction ratio (ER) almost 100\%. The ramp-up and ramp-down dynamics are quite sharp, with a typical measured temporal delay $\approx 0.14 \upmu$s, corresponding to 3-dB ER bandwidth of 7 MHz (Fig. \ref{pic:Fig4}c, also see details in Supplementary Movie 3). 
The modulation bandwidth is mainly limited by the on-off speed of the control laser, but it is still over 2 orders of magnitude faster than the state-of-the-art turnkey solitons \cite{shen2020integrated}.

In summary, we experimentally demonstrated the soliton slingshot effect and the dynamical control of counter-propagating solitons by using erbium-doped active microsphere resonators. Besides the fundamental significance in physics, the optical gain implementation in the soliton system also offers unprecedented capabilities to actively regulate the multi-soliton states, the soliton access range, the spectral spans, and the propagation chirality of the solitons, all in one device. Such realizations also pave the way for the ultrafast switch of the soliton microcombs and the tunable dual comb spectroscopy. Finally, due to its physical generality, this scheme also could be applied in other microresonator platforms, to attain soliton microcombs with broader span and richer diversity, further expanding the potential of microcomb tools for wide applications such as optical communication and sensing.

\vspace{6pt}

\noindent \textbf{Methods}\\
\begin{footnotesize}
\noindent \textbf{Theoretical analysis.} For understanding the soliton slingshot, the multimode distribution in the microsphere is displayed via electric field simulation; then deduce the coupled mode equations considering mode crossings, and then analyze the intracavity power prior to comb emergence and the threshold of four-wave mixing based on side mode excitations. In simulation, we demonstrate the comb evolution in both spectral and temporal domain by using the LLE. For understanding the chiral comb emission with controllability, the influence of scattering, the $Q$ factor and the coupling for resonances are investigated, and the mode indices are calculated. Detailed theoretical model and numerical simulations are shown in Supplementary Information Section 1.

\vspace{6pt}
\noindent \textbf{Fabrication of the Er-doped microspheres.} We fabricate the active microsphere resonators based on electrical arc discharging based thermal melting-shaping technique, which is implemented in high power fiber fusion splicers (FITEL S184; FITEL S178). To find the best performance, three specific methods are used to dope erbium ions in or on the surface of the microspheres: I. directly discharge erbium-ytterbium heavily doped fiber (EYDF) to obtain rare earth doped microspheres (diameter $>$ 300 $\upmu$m). II. We use hydrofluoric acid (HF) to etch the EYDF’s cladding but retain the core. Then we use the same method mentioned in I, the microsphere samples were prepared by using smaller discharge current and shorter discharge time (diameter $<$ 100 $\upmu$m). III. We use liquid based surface doping technology to realize Er$^{3+}$ surface doped microspheres, after arc annealing (diameter $>$ 300 $\upmu$m). The microspheres produced by this method have good doping uniformity with high quality factor up to $10^8$, and high surface Er$^{3+}$  doping concentration ($10^{19}$ level) simultaneously. Fabrication flow and characterizations are shown in Supplementary Information Section 2.

\vspace{6pt}
\noindent \textbf{Experimental details.} We check the transmissions and the $Q$ factors of our microcavities by sweeping the ECDL wavelength from 1540nm to 1560nm at a scanning speed of 1 nm/s, with fixing the ECDL power low enough avoiding ring-down. The tunable laser is connected to an oscilloscope to provide a trigger. Spectral sampling rate in this measurement is down to single kHz, which can identify one resonance well, whose half-width is in MHz level. For soliton comb generation, we detune the C-band tunable ECDL into the resonances. Meanwhile, another 980 nm diode laser can also be used for stimulating erbium gain. The maximum power of our 980 nm pumping laser is 200 mW. All light is launched in and collected by a tapered fiber with minimum diameter 1 $\upmu$m. We use a commercial frequency resolved optical gating (FROG, MesaPhotonics) to measure the soliton pulse duration and the repetition rate. It contains a motor controlled 30 cm long tunable mechanical travel, a BBO crystal to generate SHG autocorrelation, and a visible band OSA (ocean optics) for spectral analysis, with spectral resolution 0.2 nm. Measurement window 20 ps, sampling rate 10 fs/point.

\vspace{6pt}

\noindent \textbf{Acknowledgment}

\noindent The authors acknowledge support from the National Science Foundation of China (61705032, 61975025, 11825402, 12041602 and 41527805). C.W.W. is supported by the National Science Foundation of United States (1741707, 1824568 and 1810506). Y.-F.X. $\&$ Q.G. are also supported by Key R$\&$D Program of Guangdong Province (2018B030329001).

\vspace{6pt}
\noindent \textbf{Author contributions}

\noindent B.Y., Y.-F.X. and Y.R. led this project. T.T. and Z.Y. fabricated the silica based active microspheres, implemented the erbium doping and performed the characterizations. T.T., Z.Y., H.J.C., Q.T.C., and B.Y. designed the experimental scheme. T.T., Z.Y., N.A., and B.Y. built the experimental setups, performed the optical and electronic measurements. H.-J.C., Y.Y., B.Y., Q.G. and Y.-F.X. made the theoretical model and numerical simulations. All authors discussed and analyzed the results. B.Y., H.-J.C., Y.Y., Q.-T.C., T.T., Y.-F.X., and Y.R. prepared the manuscript.

\vspace{6pt}
\noindent \textbf{Competing interests} 

\noindent The authors declare no competing interests.
\end{footnotesize}

\bibliographystyle{naturesaa}

\bibliography{Reference.bib}

\begin{figure*}[hbtp]
\centering
\captionsetup{singlelinecheck=no, justification = RaggedRight, labelfont={bf},name={Extended Data Fig.}}
\setcounter{figure}{0}
\includegraphics[width=18cm]{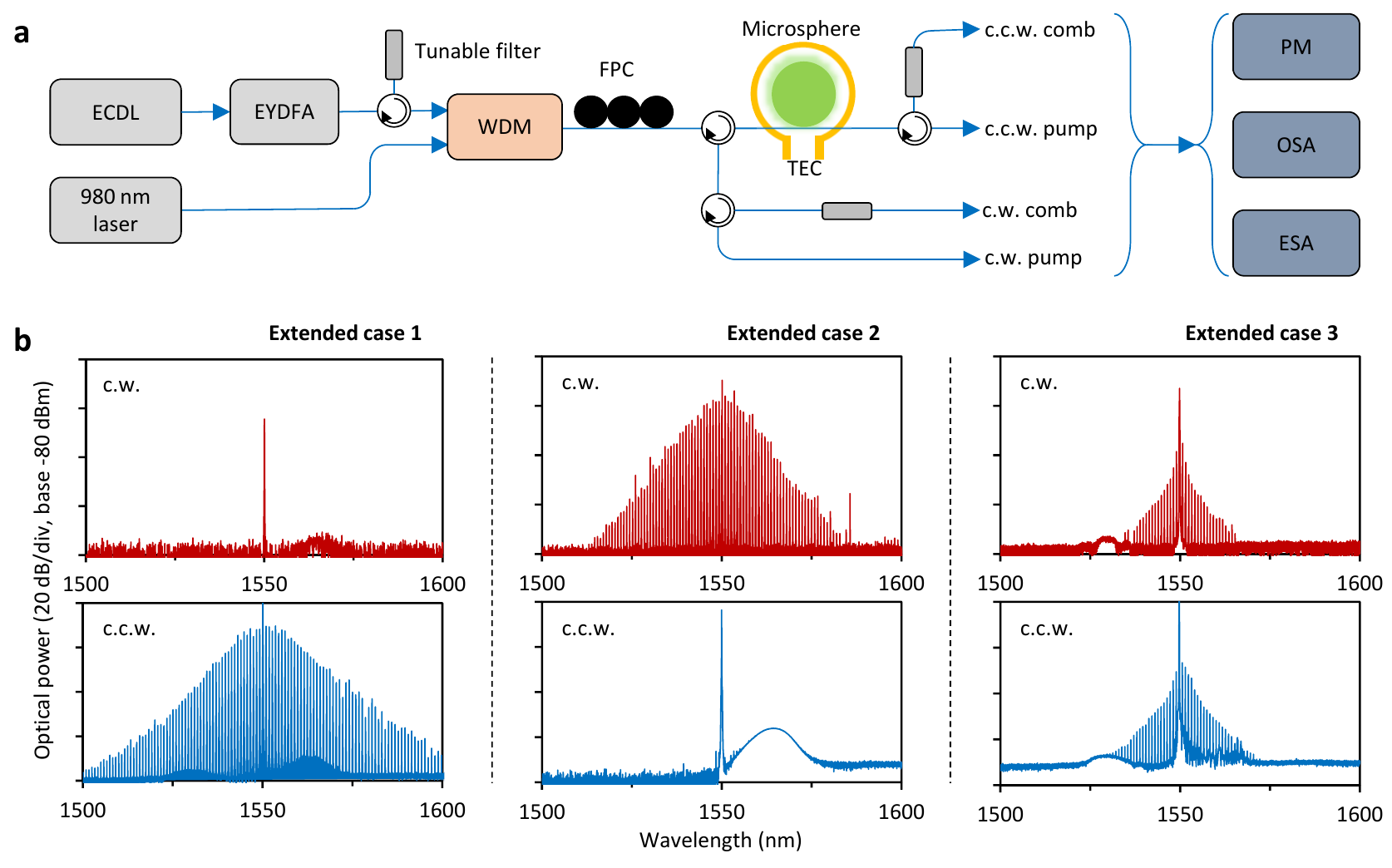}
\caption{\textbf{Experimental setup and extended cases of chiral comb generation. a,} Schematic of the experimental setup. ECDL: external cavity diode laser. EYDFA: erbium and ytterbium co-doped fiber amplifier. FBG: fiber bragg grating. WDM: wavelength division multiplexer. FPC: fiber polarization controller. TEC: temperature controller. PD: photodetectors. PM: power meter. OSA: optical spectrum analyzer. ESA: electrical spectrum analyzer. OSC: oscilloscope. \textbf{b,} Extended cases of comb generation in c.w. direction and c.c.w. direction. From left to right: comb formation in only c.c.w. direction, c.w. direction and both.}
\label{pic:ED1}
\end{figure*}

\begin{figure*}[hbtp]
\centering
\captionsetup{singlelinecheck=no, justification = RaggedRight,labelfont={bf},name={Extended Data Fig.}}
\includegraphics[width=18cm]{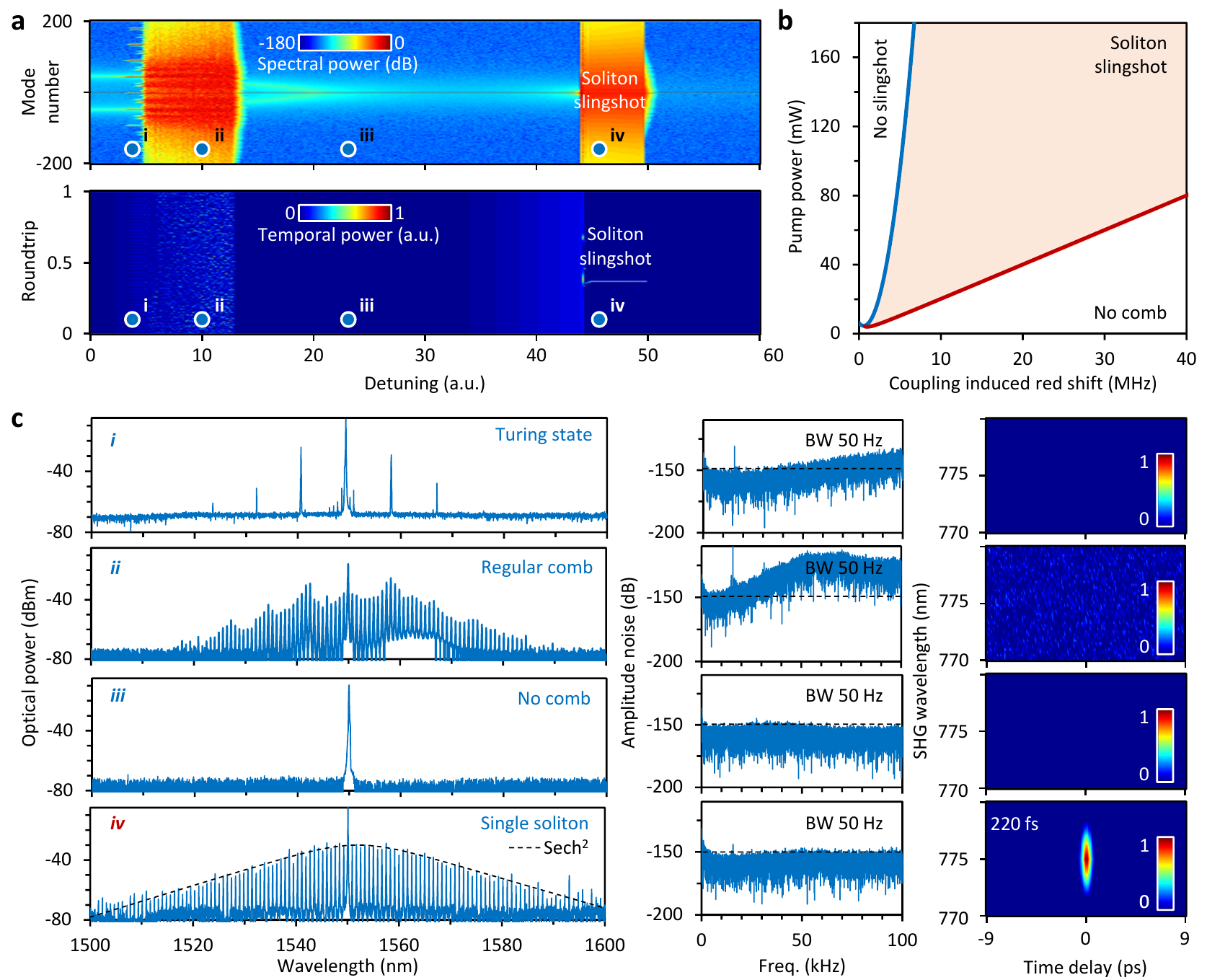}
\caption{\textbf{Evolution of single soliton slingshot. a,} Simulated comb evolution maps, in both frequency domain (top) and time domain (bottom). b, Parametric space of the comb slingshot, relying on both pump power and modal coupling based red-shift. c, Measured soliton comb evolution. Panels from left to right: optical spectra, corresponding radio-frequency amplitude noise, and time-resolved autocorrelation maps, from state i to state iv.}
\label{pic:ED2}
\end{figure*}

\begin{figure*}[hbtp]
\centering
\captionsetup{singlelinecheck=no, justification = RaggedRight, labelfont={bf},name={Extended Data Fig.}}
\includegraphics[width=18cm]{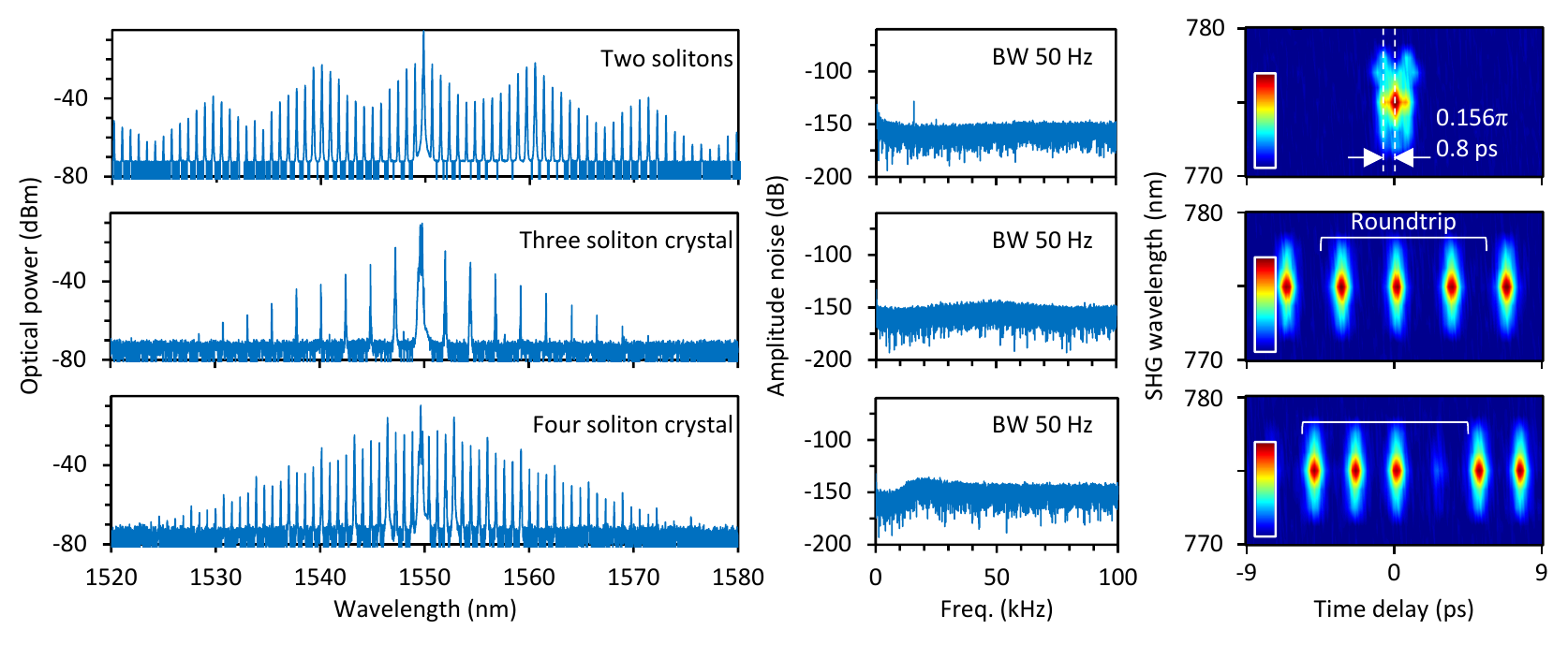}
\caption{\textbf{Diverse multi-soliton states under erbium gain assistance.} Left panels: optical spectra. Middle panels: low frequency noise spectra below 100 kHz. Right panels: frequency-resolved second-harmonic autocorrelation maps. Color bar: normalized intensity.
}
\label{pic:ED3}
\end{figure*}

\begin{figure*}[hbtp]
\centering
\captionsetup{singlelinecheck=no, justification = RaggedRight,labelfont={bf},name={Extended Data Fig.}}
\includegraphics[width=18cm]{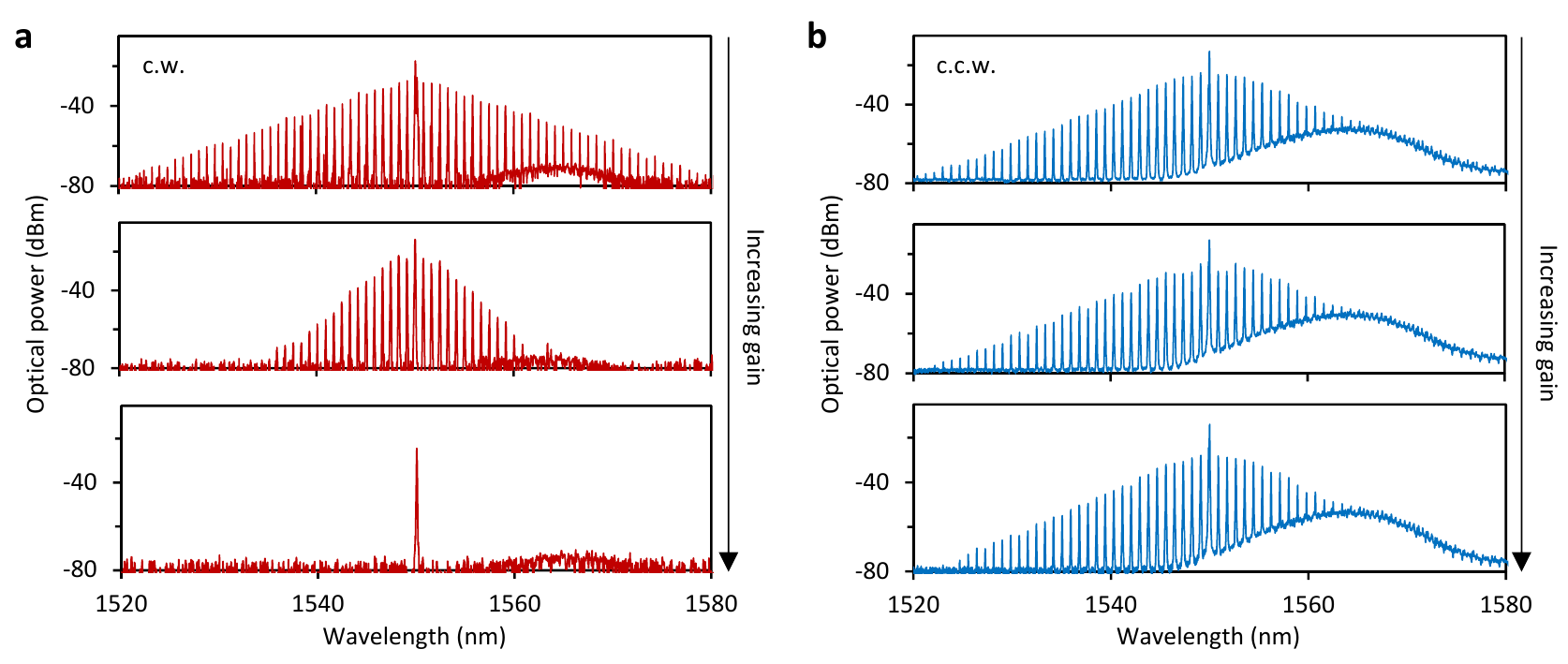}
\caption{\textbf{Output spectrum evolution in the active tuning process.} Measured comb spectra in both c.w. (a) and c.c.w. direction (b) when increasing the control laser power, corresponding to the case shown in Figure 4. Here the c.w. soliton is switched off, while the output spectrum of the c.c.w. soliton keeps almost unchanged.}
\label{pic:ED4}
\end{figure*}

\end{document}